\documentclass[letterpaper,twocolumn,10pt]{article}
\usepackage{usenix2019_v3}

\usepackage{tikz}
\usepackage{amsmath}

\usepackage{filecontents}

\RequirePackage{amsmath}[2016/11/05]
\usepackage{graphicx}
\usepackage{amssymb}
\usepackage{bm}
\usepackage{xcolor}
\usepackage[normalem]{ulem}
\usepackage{hyperref}

\begin{document}

\date{}

\title{\Large \bf Modeling infection methods of computer malware in the presence of vaccinations using epidemiological models: An analysis of real-world data}

\author{
{\rm Nir Levy}\\
Microsoft\\
nirlevy@microsoft.com
\and
{\rm Amir Rubin}\\
Microsoft,\\
Ben-Gurion University of the Negev\\
amirrub@cs.bgu.ac.il
\and
{\rm Elad Yom-Tov}\\
Microsoft Research\\
eladyt@microsoft.com
} 

\maketitle

\begin{abstract}

Computer malware and biological pathogens often use similar mechanisms of infections. For this reason, it has been suggested to model malware spread using epidemiological models developed to characterize the spread of biological pathogens. However, to date, most work examining the similarities between malware and pathogens using such methods was based on theoretical analysis and simulation.

Here we extend the classical Susceptible-Infected-Recovered (SIR) epidemiological model to describe two of the most common infection methods used by malware. We fit the proposed model to malware collected over a period of one year from a major anti-malware vendor. We show that by fitting the proposed model it is possible to identify the method of transmission used by the malware, its rate of infection, and the number of machines which will be infected unless blocked by anti-virus software. In a large sample of malware infections, the Spearman correlation between the number of actual and predicted infected machines is 0.84.

Examining cases where an anti-malware ``signature'' was transmitted to susceptible computers by the anti-virus provider, we show that the time to remove the malware will be short and independent of the number of infected computers if fewer than approximately 60\% of susceptible computers have been infected. If more computers were infected, the time to removal will be approximately 3.2 greater, and will depend on the fraction of infected computers.   

Our results show that the application of epidemiological models of infection to malware can provide anti-virus providers with information on malware spread and its potential damage. We further propose that similarities between computer malware and biological pathogens, the availability of data on the former and the dearth of data on the latter, make malware an extremely useful model for testing interventions which could later be applied to improve medicine. 

\end{abstract}

\section{Introduction}

It has been observed that biological pathogens share similarities with computer malware in their methods of infection \cite{berger2005}. These similarities have so far been limited to the application of models inspired from biology to the modeling of malware. Here we show that the parallels between the two runs deeper, and can be used to enrich our knowledge of both.

There are two primary modes of contagion in the biological world: One where the pathogen spreads through peer-to-peer (P2P) infection and the other where the pathogen infects whenever a susceptible individual visits an infected location.

Most observations on the similarity of malware to biological pathogens has focused on pathogens and malware which spreads through P2P infections. Diseases which spread through P2P contact include, for example, influenza, which is spread through airborne viruses exhaled by an infected individual. Similarly, P2P malware include, for example, malware which is spread through infected email messages such as the notorious ILOVEYOU\cite{iloveyou} and MyDoom\cite{mydoom} worms. Malware becomes activated when the email is opened by a user, which then scans the user's address book and sends itself to all the addresses therein. Another example is the recent WannaCry\cite{wannacrypt} ransomware, which used the EternalBlue \cite{eb} exploit to propagate from an infected machine.

Many food-borne biological pathogens spread when a central food processing facility is contaminated. In such cases, infection occurs only in those foods which passed through the facility. The analogous malware is one which infects computers which visit a contaminated website such as the Zeus\cite{zeus} malware, the JavaScript based drive-by cryptocurrency miner such as Trojan:JS/Miner.A\cite{miner} or the Flashback\cite{flashback} trojan, which downloads the malicious payload through the browser. We refer to this form of spreading as Central Source (CS).   

There are several additional apparent and less apparent similarities between computer malware and biological pathogens which lend the two to be modeled through similar models. 

First, the size of populations exposed to biological epidemics, which range from several hundreds of individuals to (potentially) a few billions of people or animals, is not dissimilar from the number of computers which can be potentially exposed to malware. Recent estimates \cite{gartner2017} are that 8.4 billion computers and computing devices are currently connected to the internet and are thus potentially susceptible to a malware infection. 

Though some malware, such as the notorious NotPetya\footnote{\url{https://en.wikipedia.org/wiki/2017_cyberattacks_on_Ukraine}}, is targeted at a limited number of computers, e.g., those without the patch for the EternalBlue\cite{eb} exploit, other malware infect many hundreds of millions of machines. An example of such malware is  Kovter\cite{kovter}. This is in parallel to biological pathogens, some of which can only infect a small susceptible population where others can infect most people in a population. The former usually appear when human behavior makes people susceptible to infection, and when that behavior is limited to a small group of people. For example, Kuru \cite{hoskin1969} was spread through cannibalism which was limited to a small group of people in Papua New Guinea. The latter are more common and include seasonal influenza, which can infect most people in a population \cite{waalen2010}.

Another similarity between biological pathogens and computer malware is the use of vaccines to combat their spread. Vaccines are used to inoculate a population of susceptible people or animals in order to prevent the spread of a disease within the population. Similarly, computers usually run an anti-virus software which uses {\bf signatures} of malware in order to identify potential malware files\footnote{Here we refer to software used to block malware as either anti-virus or anti-malware software interchangeably.}. These signatures are sent from a central authority or are accessed by the computers online using a cloud protection architecture. Once a signature is prepared by a security analyst and sent to computers, these computers are, with high probability, immune from the malware.

Note that vaccines to counter a pathogen are usually administered before an infection begins circulating in a population. Once a person is infected, vaccines are usually of limited efficacy (though exceptions, such as rabies, exist). In contrast, malware signatures are usually developed only after a malware is circulating in a population of computers. Computers are immune to a malware installing itself once a signature is available. Moreover, periodic scans by anti-virus software can detect and remove malware which has already been installed on a computer.  

These similarities suggest that similar tools can be used to model both pathogens and malware. Here we show that population-level modeling tools developed for pathogens can be applied to malware and that the rate of spread of malware and biological pathogens is similar. 

The most commonly used epidemiological model for infectious disease is the Susceptible-Infected-Recovered (SIR) model \cite{kermack1932}. 
This model describes the number of susceptible ($S$), infected ($I$), and recovered ($R$) individuals over time in a population, when infected by a pathogen. Individuals move from the susceptible population to the infected when the disease has spread to them, and from the infected to the recovered once they manage to overcome the infection. The SIR model is represented by a system of three ordinary differential equations. Variations in this model include the SIS model, where people who recover can be infected again, and are therefore moved directly to the susceptible population upon recovery.

The advantage of applying epidemiological modeling tools to malware is that epidemiological models have been developed for over a century and have been extensively tested. 

Perhaps even more importantly, the analogy between malware and pathogens enables the development of strategies for handling biological pathogens which can be informed by those of malware for several reasons:

First,  microscopic data for biological pathogen spread in a  population is rarely available and when it is, it necessitates complex collection efforts and is limited in scale. This is in contrast with malware, where these data are readily available to anti-virus providers. 
For example, while there are only relatively few examples of vaccinations, there are hundreds of thousands of signatures, each for a specific malware. Thus, if carefully evaluated, the spread of malware  vis-\`a-vis the development of signatures can be used to model the effects of vaccines during an epidemic or a pandemic from a biological pathogen. 

Additionally, the impact of vaccination to biological pathogens cannot be observed directly. Inoculated individual may resists the virus, but that is rarely known even by the individual. However, with computer malware, an anti-virus client will often report to the provider that it encountered and blocked a malicious file. 

We claim that the similarity among the two suggests that strategies for containment which are developed and tested on malware can then be applied to pathogens. Essentially, malware provides a model for biological pathogens which can be observed in the wild. 

Thus, we show that epidemiological models can assist in identifying the mode of transmission of malware and to estimate the size of the susceptible population. Conversely, strategies for handling of malware can inform the handling of pathogens: We show how the introduction of signatures affects the spread of malware in a population of susceptible computers, and demonstrate the importance of early vaccination to prevent wide-spread infection.


\section{Related work}
\label{sec:related}

Malware describes different types of software, including viruses and worms, that have malicious or fraudulent intent \cite{hu2009}. Some estimates are that over 350,000 new  malware and potentially unwanted application samples appear each day \cite{avtest,mlb}. 

As noted above, the idea of modeling malware spread using epidemiological models has been proposed in the past. Kephart and White \cite{kephart1992directed} applied an SIS model to a malware epidemic. In their work, the topology of the network in which the malware spreads was simulated using random graphs. They focused on computer viruses spreading through a P2P spreading mechanism. Note that the usage of random graphs, suitable for the pre-Internet era, is inaccurate for modern network topology known to be scale-free.   

Wang et al.\cite{wang2000computer} extended this analysis of virus spread to additional types of simulated networks topologies (hierarchical and clustered) and analyzed the effect of immunization on the scale of infection. The spreading mechanism of the viruses was assumed to be random (over the network topology). The analysis showed that in those types of networks, vaccination of selected nodes can help in preventing the spread of viruses, motivating further research into the selection of nodes to immune. 

Focusing on email viruses (which spread through a P2P mechanism), Gareto et al. \cite{garetto2003modeling} modeled the spread of these viruses using a stochastic model based on Markov Chains and simulated their propagation in a simple model of a small-world graph \cite{watts1998collective}. 

Berger et al. \cite{berger2005} focused on SIS models and provided theoretical support to the importance of vaccination against malware spread, rigorously showing that in a scale-free graph any virus with a positive rate of P2P spread has a positive chance of becoming an epidemic. 

All the above work used simulation in lieu of actual data on malware epidemics and network topology. Taking one step closer to real-world networks, Hu et al. \cite{hu2009} used data consisting of the location of  WiFi routers to reconstruct several router networks. Malware spread in these networks was modeled using a P2P stochastic process taking into account the router's security settings. This work, by its nature, did not consider CS spreading mechanism, and immunization of infected routers is not included in their analysis. 


Most past works assumes a homogeneous population, where the rate of infection is the same for all machines. In contrast, Feng et al. \cite{feng2013dynamical} used an SIR model with an infection rate which changes over time. This rate was a function of the number of infected machines. Heterogeneous infection rates in SIS models are also used in \cite{qu2017sis}. Here, instead of a single random-variable for any two nodes, they use a different one to model a separate infection rate for any two nodes. Their theoretical findings are validated on both synthetic and real network architectures, but without real-world malware propagation patterns.

Another SIR model which supports a more fine-grained analysis is presented in Liu et al. \cite{liu2016modeling}, where the variables controlling malware propagation and immunization  are heterogeneous. Each computer is either weakly-protected or strongly-protected. Weakly-protected computers are more likely to be infected with malware and less likely to receive a vaccination after they are infected than their strongly-protected counterparts. This work is similar to ours in that it used heterogeneous infection rates. However, whereas the infection rates in this work varied by the security level of a computer, ours depends on the malware spreading mechanism(s). A combination of the two approaches is an interesting direction for future research. 

Novel methods to disseminate the vaccine were examined in Goldenberg et al. \cite{goldenberg2005}, who analyzed simulations of malware in email networks and ways to propagate vaccines such that they will reach more susceptible machines sooner.   

Thus, our work adds several novel advances to the state of the art. First, we use a heterogeneous infection rate depending on the method of infection. Second, we fit our models to real-world data of malware propagation. Our models provide pertinent, actionable information to anti-malware vendors. Finally, we examine the effect of vaccines on malware spread in a real-world setting.



\section{Models of malware spread}
\label{sec:models}
As noted above, there are two primary models of infection used by biological pathogens and by computer malware: P2P and CS. These modes of infection can be used interchangeably by the same malware, if one mode of infection is blocked or if it exploits different vulnerabilities to spread. Thus, we propose a generalization of the SIR model \cite{kermack1932} for modeling the spread of an infectious disease in a susceptible population through both models of infection. 

The classic SIR model is represented by a system of ordinary differential equations for the number of susceptible ($S$), infected ($I$), and recovered ($R$) individuals. It accounts for a single virus or malware, and assumes peer to peer infection, that is, equal probability of being infected by any individuals regardless of his position. Formally, mean field interaction between machines can be expressed as:
\begin{equation}
	\frac{dS}{dt} = - \beta S I , \quad
    \frac{dI}{dt} = \beta S I - \gamma I, \quad
    \frac{dR}{dt} = \gamma I,
    \label{scalar_sir}
\end{equation}
where $\beta > 0$ is the infection rate and $\gamma > 0$ is the recovery rate. 

The model in Eq.(\ref{scalar_sir}), can be generalized to account for multiple malware strains and spreading mechanisms. To this end we transform to a multidimensional representation in which $S$, $I$, $R$, $\beta$, and $\gamma$ are tensors (Eq.(\ref{matrix_sir}). We further allow the two modes of infection to occur simultaneously (albeit with different infection rates) by dividing the relevant terms into two and allowing two infection rates ($ \beta_{p2p}, \beta_{CS}$). A similar model, and its reduction to account for multiple biological viruses in one population, is presented by Levy et al.\cite{levy2018modeling}. In the general case the dynamics is represented by the following system of ordinary differential equations: 

\begin{equation}
	\begin{gathered}
    \begin{aligned}
		\frac{d \mathbf{S}}{dt} &= - \mathbf{I \bm\beta^\intercal_{p2p} S} - \mathbf{\bm\beta^\intercal_{cs} S}\\
  	  	\frac{d \mathbf{I}}{dt} &= \mathbf{S \bm\beta^\intercal_{p2p} I + S \bm\beta^\intercal_{cs} - \bm\gamma^\intercal I} \\
  	  	\frac{d \mathbf{R}}{dt} &= \mathbf{\bm\gamma^\intercal I} \\
 	  	\label{matrix_sir}
	\end{aligned}
	\end{gathered}
\end{equation}

where boldface represents a tensor, and  $\bm\beta_{p2p}$ and $\bm\beta_{cs}$ are probability tensors representing the probability to be infected from a peer machine or from a central source respectively.

The main difference between the mode of propagation of P2P and that of CS (quantified in the second equation above) is in the factors determining the number of infected machines (or individuals) in the next time step. In P2P propagation this is dependent both on the number of susceptible machines and the number of infected machines at the current time, whereas in CS propagation it is only predicated on the number of currently susceptible machines.  

In this paper we focus on a special dynamic evolution case, where a system is infected with multiple malware strains that do not mutate, and where each malware propagates and infects additional machines either through peer-to-peer interactions (for example, through infected email messages) or from a central source (e.g., where visiting a specific website causes a malware to be downloaded to the computer). Hence, all tensors in Eq.(\ref{matrix_sir}) are diagonal and we allow either $\bm\beta_{p2p}$ or $\bm\beta_{cs}$ to be non-zero. 

We assume that, in contrast to biological viruses, malware will eradicated from a machine only when an appropriate security patch will be installed, and once it is installed recovery is immediate. This behaviour is similar to SIS-like dynamics, which characterize malware, see Berger et. al. \cite{berger2005}. 

The infected population of computers is important for two reasons. First, it represents the population which might suffer from the harms associated with the malware (for example, it will devote CPU cycles to mine bitcoin when the machine's owner did not intend to do so). In P2P propagation it can also infect other machines. An infected machine will continue to infect others either until it receives the relevant signature or if it is disconnected from the network or turned off. For these reasons, $\bm\gamma$ is affected by multiple factors, and should not be set to zero even if a signature is not available.

In general, the model in Eq.(\ref{matrix_sir}) can account for mutations as well as hybrid infection vectors. These are modeled by removing the restriction on all tensors. In such cases, non-diagonal elements can represent mutations. A combination of non-zero $\bm\beta_{p2p}$ and $\bm\beta_{cs}$ can represent hybrid propagation and infection vectors of malware. 

\section{Methods}
\label{sec:methods}
Henceforth we use the following notation: The time-series representing the number of machines first infected with a given malware on day $t$ is denoted by $x_{t}$, where $t=0$ is the first day that this malware appeared on any machine. The number of infected machines at day $t$ according to the models is denoted by $x_{t}^M$.

\subsection{Data}
Our dataset contains telemetry reports from Microsoft Defender Anti-Virus
collected between April 1st, 2017 to April 1st, 2018. These reports are produced by end-point machines when a file scan is performed by the anti-virus software. Each telemetry report contains metadata regarding the file, including a unique identifier which is computed based on the file's content (as identified by the SHA-1 hash of the file), as well as the time of the scan, a unique identifier of the machine,  the result of the scan performed and the time that the file first appeared on the machine.

We note that it is possible for a clean file to be scanned (for instance if downloaded from the web, or if the file is a portable-executable file) and the results of this scan to be in the reported data.

Our analysis contains all the report of files which were classified as malware sometime during the time frame used. In our analysis we focused on files which were seen on at least 200 machines.

We modelled the time series of each malware through the number of machines per day on which the malware was first seen on (regardless of it being detected as malware).

 Using these data, we found the first time that each malware was marked as such, indicating that an appropriate signature was developed for it and that this signature was available to the anti-virus software on machines. We refer to this time as the {\bf Vaccination Time}.

\subsection{Fitting models to data}
\label{sec:fitting}

The P2P model and the CS model are each parametrized by three parameters: The number of individuals infected at time zero (I(0)); the rate of recovery ($\gamma$); and the Basic Reproduction Number ($R_0$), which is equal to $\beta / \gamma$ (either $\bm\beta_{p2p}$ or $\bm\beta_{cs}$, depending on the model). Following \cite{levy2018modeling}, we generated a dictionary of model functions by creating instantiations of the number of infected individuals over time for each valid solution of the model equations in a 10-point grid with the parameter values shown in Table \ref{tbl:dictionaryParameters}. 

As noted above, in this work we focus on the case where malware strains do not mutate, and each infects other machines either through P2P interactions or from a central source. Thus, two models, each describing a different mode of propagation, are fit to every malware. 

The models are fit to the time series of a malware (the number of infected machines over time) by finding the dictionary function with the highest cross-correlation with the malware time series. The cross-correlation ($c(\tau)$) is defined \cite{rabiner1975} as: 
\[
c(\tau) = \sum_{t = T_{Min}}^{T_{Max}} d_{(t-\tau)} x_{t} 
\]

where $\tau \in [0,1,\cdots, 25]$ to reduce the effect of noise, $d_t$ is the dictionary function and $x_t$ is the time series of the malware, which is assumed to have values between $T_{Min}$ and $T_{Max}$.

The use of cross-correlation allows a phase difference (time shift) between dictionary time series and the malware time series. The dictionary function from each of the propagation methods (CS and P2P) which reached the highest correlation at any delay is chosen.

After fitting each of the models, the model (CS or P2P) with the highest cross-correlation is chosen as the model which best describes the malware. The output of this stage is $x_{t}^M$.

\begin{table*}[th]
     \centering
    	\begin{tabular}{|l|c|c|}
        \hline
        \textbf{Parameter} & \textbf{Range} & \textbf{Spacing} \\
	    \hline
        The number of individuals infected at time zero ($I(0)$) & $(10^0, 10^7)$ &  Logarithmic \\
        Basic reproduction number ($R_0$) & $(0.7, 5)$ & Linear \\
        Rate of recovery ($\gamma$) & $10^{-6}, 10^{-2}$ & Logarithmic \\
        \hline
        \hline
    \end{tabular}
    \caption{Dictionary parameters}
    \label{tbl:dictionaryParameters}
\end{table*}

It is assumed that one of the models will better fit a given malware because it describes its actual mode of propagation. To measure the accuracy of this assumption, 30 malware instances were labeled by a security expert as to their method of propagation. This was compared to the method of propagation inferred by the best fitting model.

\subsection{Predicting the size of the susceptible \\ population}
\label{sec:prediction}

The susceptible population was estimated through the following process: First, the P2P and CS models were fit using only the first 30 days since a malware first appeared. The best fitting model was chosen. 

Then, $x_t$ was scaled by fitting a first-order polynomial from it to $x_{t}^M$ for $t=0,...,30$. Thus, $x_i \approx \alpha \cdot x_{t}^M + \beta$. The model was fit to minimize the sum-of-squares of the residual.

The total number of susceptible individuals was computed as the cumulative sum of the of $x_{t}^M$, transformed using the linear model: $\alpha \cdot \sum_{t=0}^\infty x_{t}^M + \beta$. 

The model was validated by examining malware instances where an anti-virus signature was applied to the malware at least 6 months after the peak of the infection and its effect was therefore small. For those we compared the total number of machines that were infected by the malware with the predicted number of susceptible machines. 

\subsection{Measuring the effect of vaccination}

Once a malware is identified, an anti-virus ``signature'' is developed, which typifies the malware. This signature acts as a vaccine, enabling computers to identify and quarantine the malware. These signatures are either available online, for anti-virus software on client machines to access them, or are transmitted to these clients for offline use.

In this section we seek to measure the correlation between the time that the vaccine is made available vis-\`a-vis the spread of the malware, and the time it takes for the malware to stop spreading. 

Our measure of malware spread is the time until 99\% of all machines reporting to have been infected by the malware are infected. We refer to this as the {\bf Time to Termination}, denoted by $T_{Term}$. We modeled $T_{Term}$ as a function of the following attributes: 

\begin{enumerate}
    \item The number of estimated susceptible machines (as calculated in Section \ref{sec:prediction}).
    \item The number of infected machines until the time that the vaccine is provided.
    \item The model parameter $\beta$ of the best-fitting model (either $\bm\beta_{p2p}$ or $\bm\beta_{cs}$).
    \item The fit of the models (P2P and CS): The maximal cross-correlation (see \ref{sec:fitting}) of each model.
\end{enumerate}

Model fits and the estimates for the number of infected machines were computed using the first 30 days of data from the first time that the malware was observed on any machine or, if the vaccine was disseminated before, using the data until the date of vaccine dissemination.

The ability of the above attributes to explain $T_{Term}$ was evaluated in two ways: First, we trained a regression tree model and evaluated its performance using 10-fold cross validation. Second, we created a Cox proportional hazard model. In both cases the predicted size of the susceptible population and the number of infected machines at vaccination time were log-transformed because of their skewed distribution.

\section{Results}
\label{sec:results}
\subsection{Fitting models to data}

As noted in Section \ref{sec:fitting}, 30 files were analyzed by a security analyst as to their method of propagation. The method cited by the analyst (CS or P2P) was compared to that inferred from the model with the better fit to the data.

Among the 30 files examined, the method of 12 could not be deduced. The method of propagation of the remaining 18 was compared to the method of propagation inferred by the best fitting model (P2P or CS). The Kappa statistic \cite{mchugh2012} between the two was $0.48$, indicating a medium-level agreement between the method of propagation as inferred from the temporal profile of the malware and the actual method of propagation. All cases of P2P-propagating malware were correctly identified, while almost half (5 of 12) CS malware were not. 

Interestingly, among the 5 CS malware that were identified as P2P, the transmission mechanism of two malware files was predominantly CS, but they also had a P2P transmission mechanism, and hence could be considered a hybrid. If these two malware are excluded, the Kappa rises to $0.64$. This finding motivates further research on the option of a hybrid infection mechanism.

\subsection{Predicting the susceptible population}

\begin{figure}
\begin{center}
  \includegraphics[width=\linewidth]{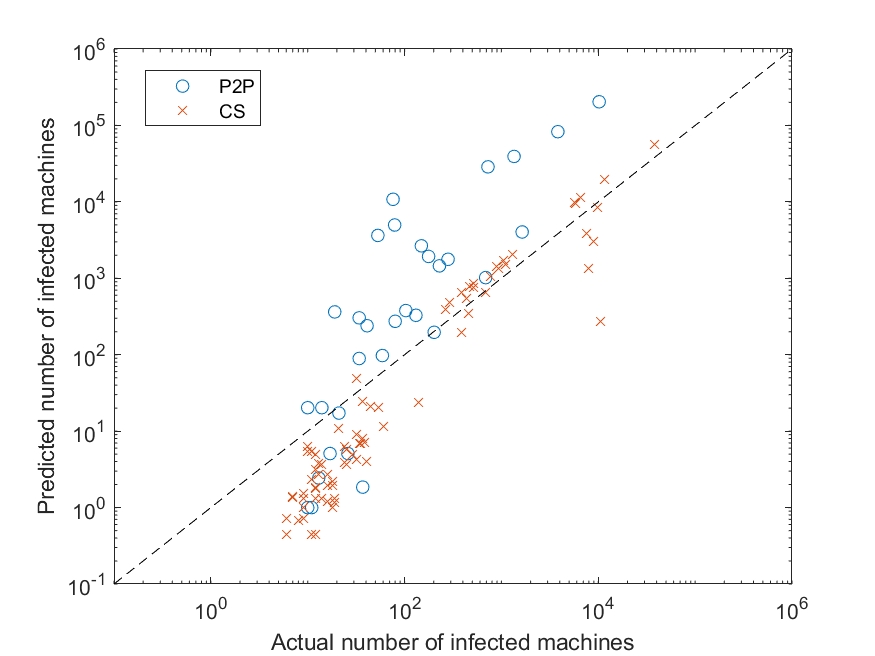}
\end{center}
  
  \caption{The predicted number of infected machines versus the actual number of infected machines. The axes are both log-scaled. The dotted line shows equal values of predicted and actual machines.}
  \label{fig:predicted_susceptible}
\end{figure}

Figure \ref{fig:predicted_susceptible} shows the plot of the predicted number of susceptible machines and their actual number, for 130 cases where the anti-virus was not applied to the malware, or was applied after most machines were infected. 

Of the 130 files, 31 were matched to a P2P model and 99 to a CS model. The average correlation between the best fitting instantiation of the SIR model and the data was 0.30. The average fit to the CS model was, on average, better (0.35), compared to the fit of the P2P model (0.16).  

As Figure \ref{fig:predicted_susceptible} shows, there is strong correlation between the predicted and actual number of infected machines (Spearman $\rho=0.84$, $P<10^{-10}$). 

Note that the estimated number of  machines is often higher than the actual number for P2P infections. This is likely due to the (typically) lower slope of the number of infected individuals at the beginning of the epidemic in P2P models, which makes the curve fitting of the linear model (Section \ref{sec:prediction}) noisier.

\subsection{Measuring the effect of vaccination}

The effect of the introduction of a vaccine was measured on 188 malware instances. 

\begin{figure}
\centering
  \includegraphics[width=0.8\linewidth]{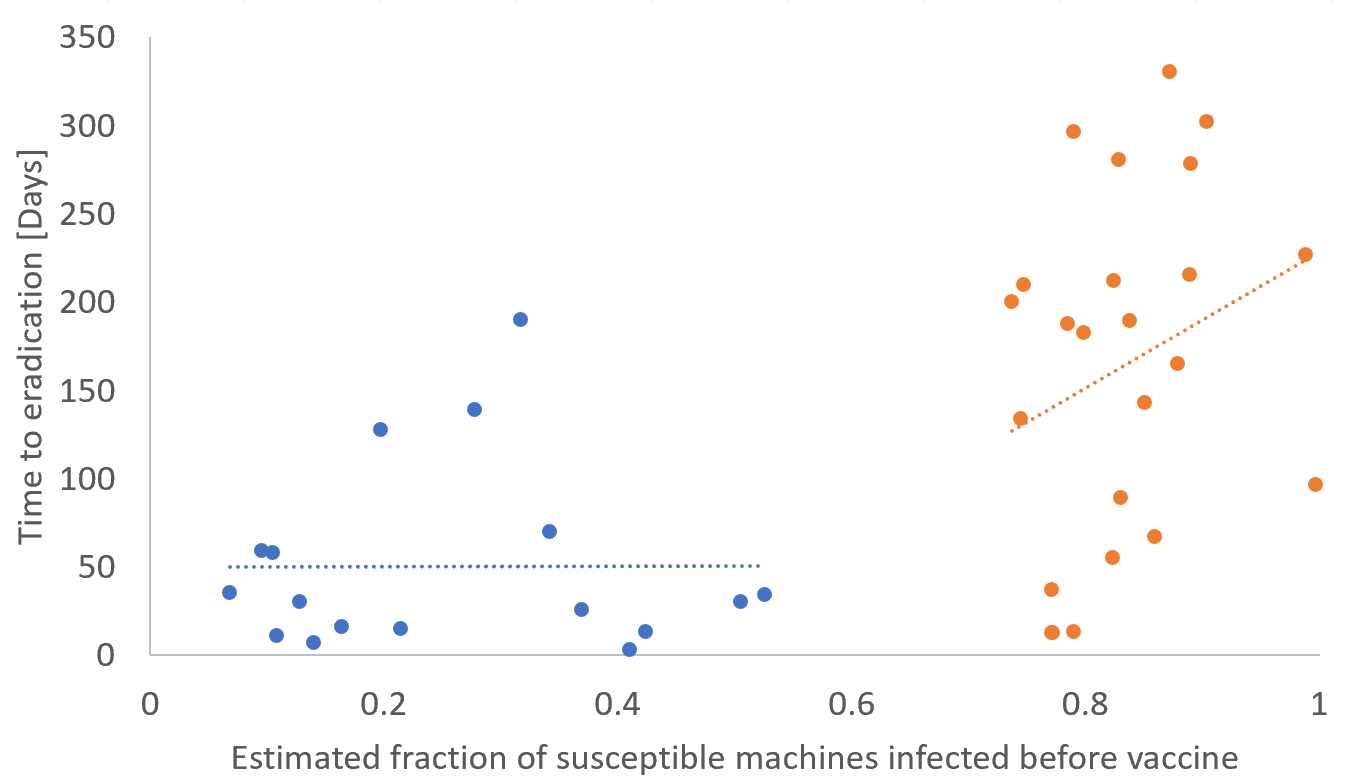}
  \caption{Time to eradication as a function of the estimated fraction of susceptible machines that were infected at the time of vaccination. Blue dots denote malware with an estimated fraction smaller than 0.6 whereas orange dots have a fraction greater than 0.6. Lines denote the fit of a linear model.  }
  \label{fig:timeToEradication}
\end{figure}

\begin{table*}[h]
     \centering
    	\begin{tabular}{|l|c|c|}
        \hline
        \textbf{Parameter} & \textbf{Hazard} (S.E.) & \textbf{p-value} \\
	    \hline
P2P Model fit & 4.36 (0.96) & N.S. \\
CS  Model fit & 13.69 (0.56) & $2.6\cdot10^{-6}$ \\
Predicted susceptible population & 1.08 (0.10) & N.S. \\
Number of infected machines at vaccination time & 0.67 (0.14) & 0.005 \\
Model parameter & 0.00 (17.01) & N.S. \\
        \hline
        \hline
    \end{tabular}
    \caption{Hazard model coefficients of the Time to Termination ($T_{Tot}$). N.S. refers to a parameter which is not statistically significantly associated with the Time to Termination. }
    \label{tbl:hazard}
\end{table*}

Figure \ref{fig:timeToEradication} shows the time 
to eradication as a function of the estimated fraction of susceptible machines that were infected at the time of vaccination. In the figure, malware instances are divided to two populations according to the fraction of infected machines at the time of vaccination. For the population with a fraction smaller than 0.6 the average time to eradication is 51 days. A rank regression model shows no correlation between the fraction of susceptible infected machines and the time to eradication ($R^2=0.01$, $P=0.64$). This is in opposition to the population where the fraction of infection is greater than 0.6. For this population the average time to eradication is 164 days, and a rank regression model achieves $R^2=0.14$ ($P=0.07$). Thus, the time to eradication is approximately 3.2 times greater when the fraction of infection is greater than 0.6.

Thus, when an infection has not been able to infect a significant part of the susceptible population, the time to eradication is not affected by the phase of infection where vaccination was administered and the time to eradication is short. However, if much of the susceptible population is infected, time to eradication is dependent on when the vaccine was administered and, on average, eradication will take much longer. 

The regression tree model reached a Spreaman correlation of $\rho=0.35$ ($p=6\cdot10^{-7}$). This indicates that the Time to Termination can be predicted with a significant accuracy. 

Attribute importance was estimated using Shapley Additive Explanations \cite{Lundberg2017}. The most important attributes were CS model fit and the number of infected machines at vaccination time. We posit that these two were the most influential variables because most malware were CS, and because in CS infections the number of infected machines in the next time step is solely determined by the number of infected machines in the current time step. Hence, these two parameters interact to determine the Time to Termination. The fit of the P2P model was the 3rd most important attribute, and was negatively correlated with Time to Termination. 

The hazard model is shown in Table \ref{tbl:hazard}. It is noted that a high hazard rate means that malware are more likely to be terminated. The hazard model found two statistically significant variables: The quality of fit to the CS model and the number of infected individuals at vaccination time. The first shows that a good CS model fit increases the chances that the vaccination will terminate the infection, possibly indicating that CS malware are easier to terminate using vaccines. The number of infected machines at vaccination time has a hazard ratio of 0.67, indicating that it is harder to stop an infection if more machines are already infected.

\section{Discussion}
\label{sec:discussion}
Spreading mechanisms of computer malware and of biological pathogens share significant underlying similarities. In this paper we discuss two spreading mechanisms, Peer-to-Peer (P2P) and Central Source (CS), which are common in both malware and biological pathogens. Our approach starts with a model that can describe the number of infected machines or individuals when a pathogen or malware uses each of the spreading mechanisms. We also offer a hybrid model to describe malware types that use both mechanisms simultaneously (though we do not demonstrate its application in this paper). 


Our findings demonstrate that it is possible to infer the spreading mechanism of a malware from the temporal profile of the model fit to it. This fit also provides an estimate of the susceptible population soon after the malware begins spreading (Figure \ref{fig:predicted_susceptible}).

Our models suggest that vaccines (malware signatures) are most effective for reducing the spread of CS malware. Moreover, once a malware is installed on a significant fraction of potentially infected machines, it is much more difficult to remove it (see Figure \ref{fig:timeToEradication}). Interestingly, this behavior demonstrates a phase transition whereby the time to remove a malware is independent of the percentage of susceptible machines infected until around 60\% of potential. Such behavior has been observed in simulations of biological infections \cite{zaman2008}.

The use of a model applied to observed infection dynamics enables us to estimate the spreading mechanism of the malware and from it, hidden variables typifying the malware and its environment. These include the probability of infection ($\beta$) and the number of potentially vulnerable machines -- the susceptible population. These two parameters are important information for an anti-virus vendor, as they allow an assessment of the speed with which the vendor needs to respond to a new threat: The probability of infection indicates how quickly a malware will spread while the number of susceptible machines shows how many machines will be infected if no action is taken. 

Thus, fitting the epidemiological model can help  anti-virus vendors to achieve two important goals. First, the model which obtains the best fit hints to the way that it spreads, suggesting ways to counter it. Second, the parameters of the model can help prioritize the response to malware because it informs of its reach and speed of propagation. We note that for prioritization, an important third factor, the harm done by each malware, should be taken into account. Our models do not provide such assessment. 

The behavioural likeness of viruses of the cyber domain and those of the biological nature can help in understanding of various processes in one domain from observations in the other. An example for such cross domain postulation is the effect of vaccination, which is easily tracked in the cyber domain and hard to follow in the biological one. Our finding that for malware, the timing and nature of spreading relative to the timing of vaccination influences the effectiveness of vaccination, can be extrapolated to biology. Our finding that there is a threshold effect, whereby the time to eradication is independent until approximately 60\% of the susceptible population is infected means that public health authorities should act quickly to suppress an infection, especially if it is estimated that the infection is close to the critical threshold.  

Future work will examine other parallels between malware and biological pathogens, and the knowledge that can be gleaned about one from observing the other. For example, the effect of population heterogeneity and of geography and mobility are known to affect the pattern of spread of biological pathogens \cite{merler2009,balcan2009,oren2018}. The concept of geographical proximity (both physical and logical) also exists in computer networks and its effect on malware spreading has begun to be examined \cite{chen2005,hu2009}. Future work will determine if the epidemiological analogy is indeed useful to facilitate monitoring and blocking of malware which exploits proximity-related weaknesses.

Another avenue for future investigation will determine the limits of the analogy. Thus, just as it is important to determine when knowledge on malware can inform health authorities and vice versa, it is crucial to understand when it does not. We are examining areas where the proposed analogies break.

Nevertheless, we believe that the analogies between malware and pathogens are useful and can help both health authorities and anti-virus providers more insights to combating these threats.

{\bf Acknowledgements:} The authors would like to thank Lev Muchnik for enlightening discussions and comments.

\bibliographystyle{plain}
\bibliography{bibliography}

\end{document}